\documentclass{article}

    \PassOptionsToPackage{numbers, compress}{natbib}

 \usepackage[main, final]{neurips_2025}
\workshoptitle{AI for Music}

\usepackage[utf8]{inputenc} 
\usepackage[T1]{fontenc}    
\usepackage{hyperref}       
\usepackage{url}            
\usepackage{booktabs}       
\usepackage{amsfonts}       
\usepackage{nicefrac}       
\usepackage{microtype}      
\usepackage{xcolor}         
\usepackage{verbatim}
\usepackage{graphicx}
\usepackage{subfigure}
\usepackage{amsmath}
\usepackage{enumitem}
\usepackage{wrapfig}
\usepackage{float}

\title{Advancing Multi-Instrument Music Transcription: Results from the 2025 AMT Challenge}

\author{
\normalfont
  Ojas Chaturvedi, Kayshav Bhardwaj, Tanay Gondil,
  Benjamin Shiue-Hal Chou, \\
  \{ochaturv,
  bhardw43, tgondil, chou150\}@purdue.edu \\
    Kristen Yeon-Ji Yun, Yung-Hsiang Lu, Yujia Yan, Sungkyun Chang \\
  \{yun98, yunglu\}@purdue.edu,    yujia.yan@rochester.edu, sungkyun.chang@qmul.ac.uk 
}

\begin{document}

\maketitle

\begin{abstract}


    This paper presents the results of the 2025 Automatic Music Transcription (AMT) Challenge, an online competition to benchmark progress in multi-instrument transcription. Eight teams submitted valid solutions; two outperformed the baseline MT3 model. The results highlight both advances in transcription accuracy and the remaining difficulties in handling polyphony and timbre variation. We conclude with directions for future challenges: broader genre coverage and stronger emphasis on instrument detection.

\end{abstract}


\section{Introduction}

Automatic Music Transcription (AMT) converts audio signals into symbolic representations of music, such as sheet music or MIDI (Musical Instrument Digital Interface) format. Compared to other fields of artificial intelligence (such as natural language processing and computer vision), the progress of AMT has been slower~\cite{benetos_automatic_2019}. Several factors contribute to this gap: (1) lack of large openly available datasets for training and evaluation, (2) absence of commonly adopted benchmarks for comparing different methods, (3) limited incentives to attract researchers compared to other fields in AI, (4) lack of a common platform where multiple AMT solutions can be evaluated., and (5) the inherent complexity of music signals. A single instrument can produce wide variations in pitch, articulation, dynamics, and playing techniques. Note onsets/offsets themselves can be ambiguous; e.g.,  piano notes may be defined acoustically (sounding strings) or by key press and release with or without pedals. In this work, we adopt the \texttt{mir\_eval} convention~\cite{raffel_mir_eval_2014}, where onsets are defined by the reference MIDI start time and offsets by the later of 50 ms or 20\% of note duration.
This paper presents the results of the 2025 Automatic Music Transcription Challenge, designed to advance AMT technology by fostering benchmarking, comparison, and community engagement.

\begin{figure*}[h]
    \centering
    \subfigure[]
    {\includegraphics[width=2.7in]{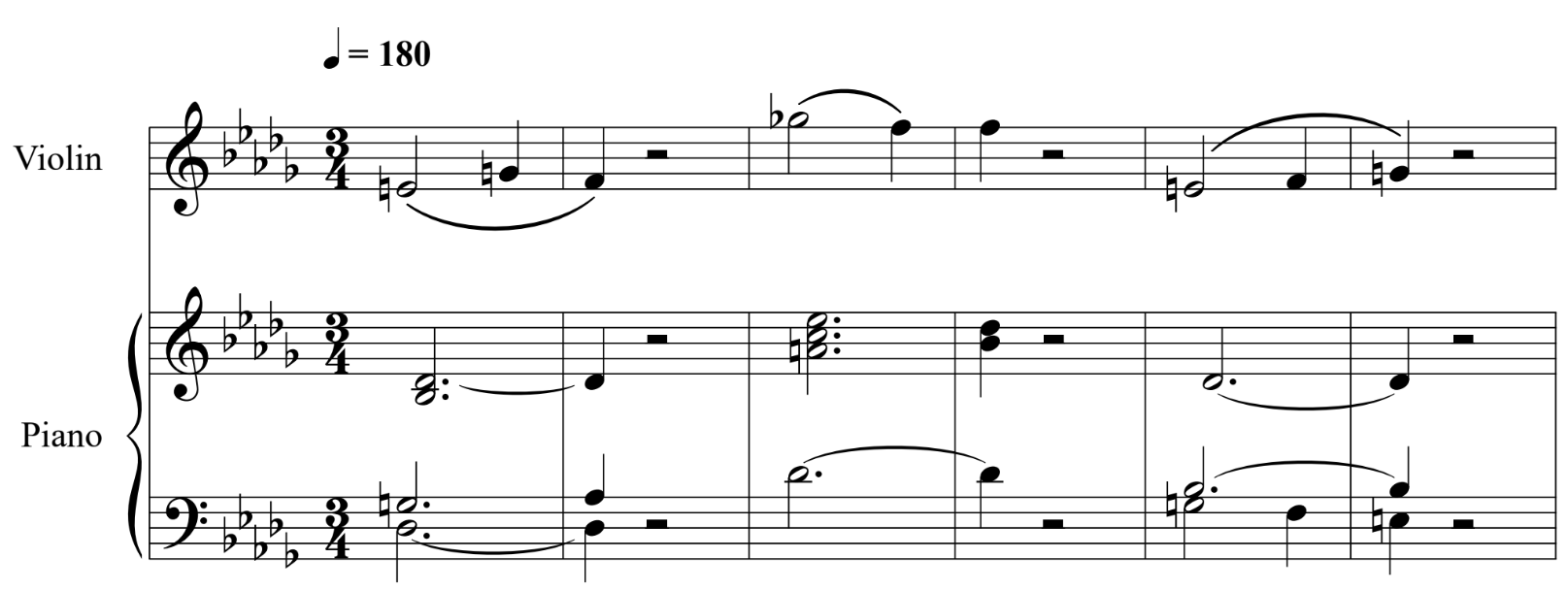}}
    \subfigure[]
    {\includegraphics[width=2.7in]{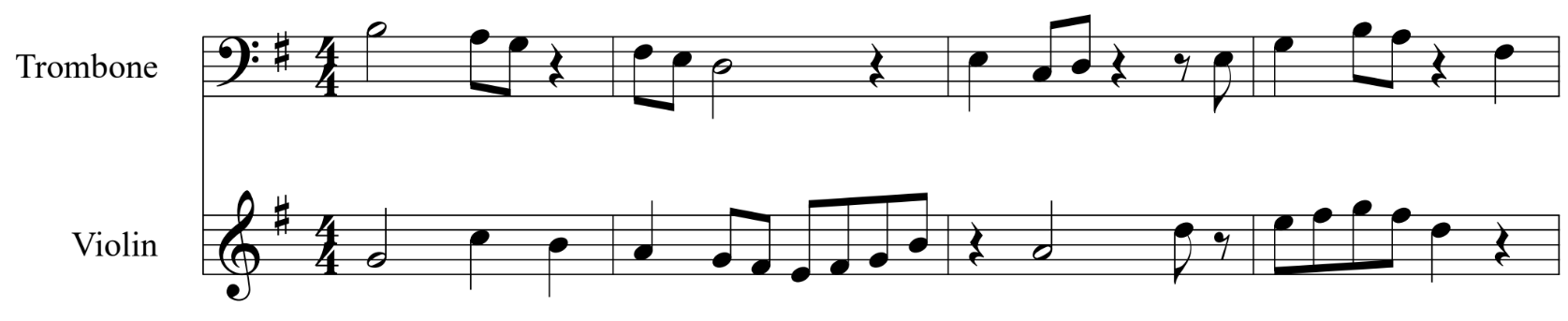}}
    \caption{Excerpts of the sample music available to the participants.
    }
    \label{figure:music}
\end{figure*}

This challenge is different from previous competitions in several ways:
(1) This challenge offers a cloud-based grading system that updates the leaderboard on a daily basis. Such timely information encourages participants to improve their methods progressively.
(2) The evaluation metrics include testing for multiple factors, as well as taking into account tolerances in pitch.
(3) To avoid overfitting to existing datasets, the challenge introduced a newly composed test set comprising 76 short pieces ($\approx$20 seconds each) across eight instruments;
up to three instruments may be present in each piece. Figure~\ref{figure:music} shows two excerpts of the released music. Each piece includes three files: (1) a PDF file of the sheet music, (2) an MP3 of synthesized audio, and (3) a MIDI file.




\section{Previous Transcription Competitions}

Music transcription has been evaluated through a variety of  competitions, each designed to address specific aspects of the task and to drive progress. Since its inception in 2005, the Music Information Retrieval Evaluation eXchange (MIREX)~\cite{noauthor_mirex_2025} has served as a pivotal benchmark, initially focusing on foundational tasks such as onset detection, tempo and beat tracking, melody extraction, key detection, and chord estimation, with submissions typically targeting one component of the broader transcription pipeline. Over time, challenges evolved toward instrument-specific, polyphonic settings, notably the Drum Transcription challenge~\cite{noauthor_drum_2021} and the Polyphonic Piano Transcription challenge~\cite{noauthor_polyphonic_2024}. The former emphasized detecting bass drum, snare drum, and hi-hat in polyphonic mixes;  the latter (introduced in 2024) required systems to convert solo piano audio into MIDI by capturing onsets, offsets, pitch, and velocity, leveraging the MAESTRO v3.0.0 dataset for training and evaluation.

\section{Evaluation Methodology}


\begin{figure}[ht]
    \centering
    \includegraphics[width=0.99\textwidth]{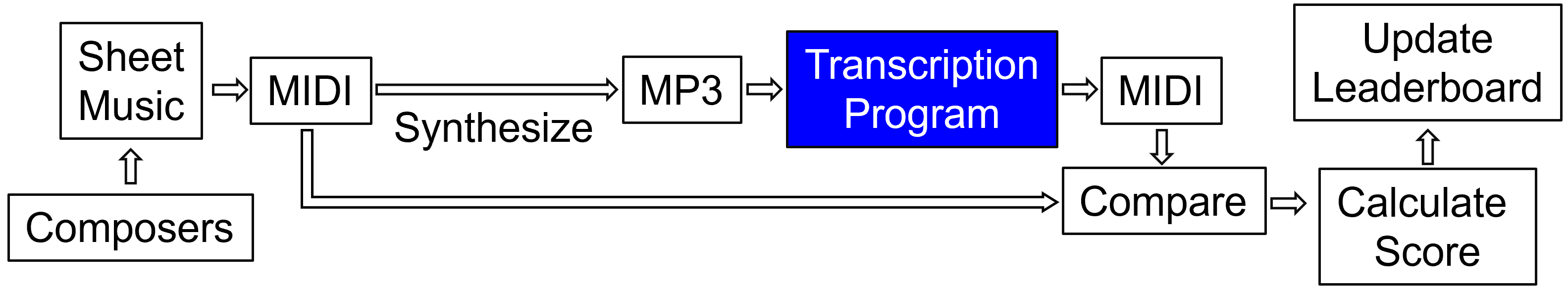}
    \caption{End-to-end data flow of the transcription challenge.}
    \label{figure:dataflow}
\end{figure}

\begin{wrapfigure}{r}{0.26\textwidth}
    \centering
    \includegraphics[width=0.29\textwidth]{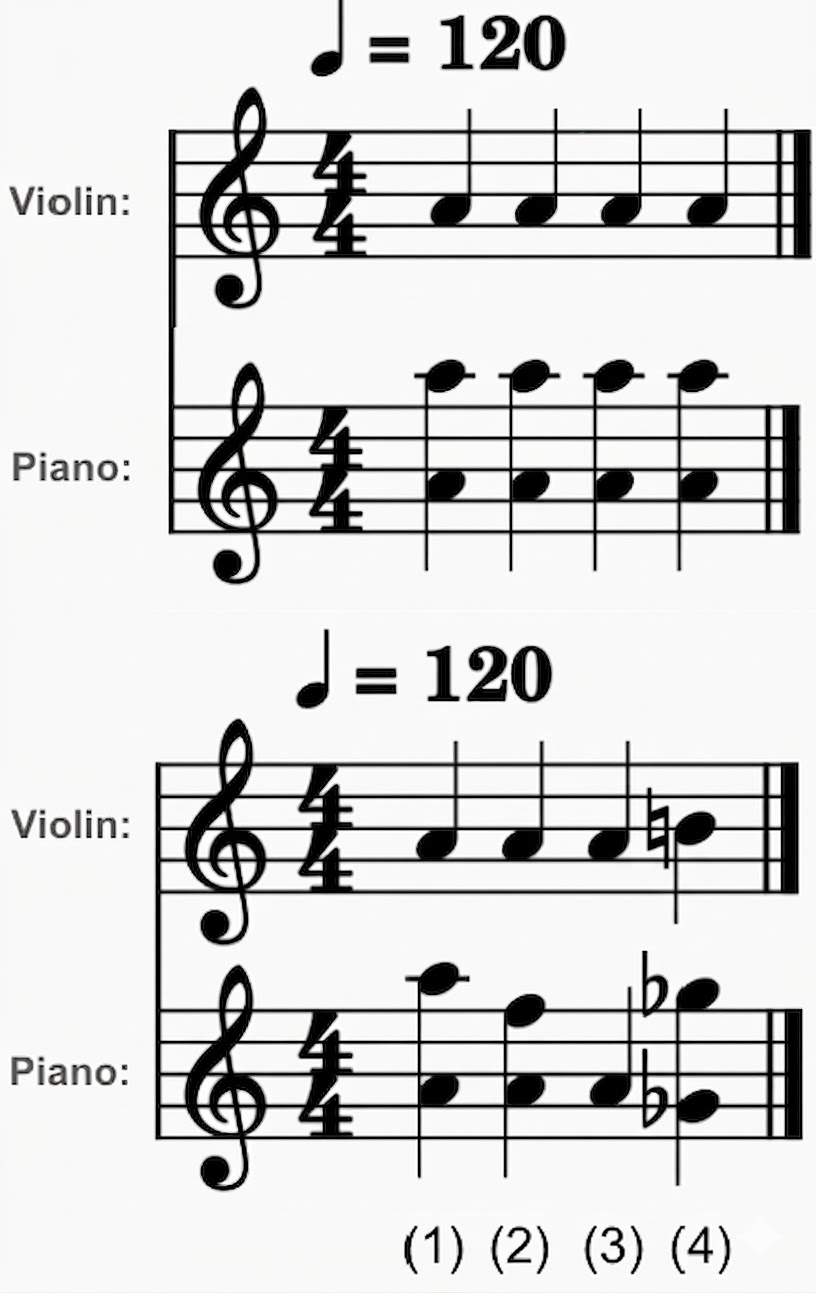}
    \caption{Evaluation Method. Top: Reference. Bottom: Estimated notes.}
    \label{fig:placeholder}
    \vspace{-10pt}
\end{wrapfigure}
To ensure a standardized and reproducible benchmark, we developed a cloud-based evaluation pipeline that automates the transition from symbolic music to acoustic signals and back to symbolic evaluation. Figure~\ref{figure:dataflow} illustrates this end-to-end process.

The pipeline begins by rendering original MIDI scores into MP3 audio; the specifics of this synthesis process are detailed in Section~\ref{subsection:testdata}. Once a participant's model generates an estimated MIDI file, it is taken in by the grading module for automated comparison against the reference ground truth as described in Section~\ref{subsec:scoring}.

\subsection{Evaluation System}
\label{subsec:scoring}

Two MIDI files are compared to calculate the transcription program's score. The evaluation metrics include precision, recall, F1 score, and overlap ratio. The precision and recall of the reference and estimated MIDI are computed using the mir$\_$eval library~\cite{raffel_mir_eval_2014}. 
Multi-instrument Note Onset F1 (a.k.a. Multi Onset F1) is a prediction that is correct only if instrument/program, pitch, and onset (±50 ms) match the reference, and the piece-wise score is macro-averaged across instruments (run mir$\_$eval per instrument, then average). For each piece, we compute per-instrument F1 scores and take an unweighted mean across instruments; then report the unweighted mean across pieces. Overlap is calculated through onset and offset, i.e., the timing of a music note's beginning and ending, by computing the intersection over union between transcription and ground truth.

Consider Figure~\ref{fig:placeholder}. The top is the reference (correct).
The bottom is the estimated (transcription).
The excerpts are divided into four groups. Each group is one quarter note long, and each group has a slight variation from the reference. Group (1) has the reference and estimated tracks as identical and receives an F1 score of 1.0. Group (2) has the top note in the piano part as incorrect. It correctly finds two notes, but misses a correct one, which is counted as a false negative. An incorrectly predicted note is counted as a false positive. Using the formulas, precision and recall are both 0.667, leading to group (2) and receiving an F1 score of 0.667. Group (3) has the top note of the piano part as completely missing. Therefore, it has two true positives for the correct note, one false negative for the missing note, and no false positives. Precision is 1.00, and recall is 0.667. Group (3)  receives an F1 score of 0.800. Group (4) has all three notes from both instruments incorrect. It has three false negatives for the missing true positive notes and three false positives for the wrong notes, with no true positives, causing precision and recall to be 0.  Group (4) receives an F1 score of 0.000. The overall F1 score of the entire excerpt is 0.609.

\subsection{Test Data}
\label{subsection:testdata}

To fairly benchmark transcription models, the challenge used a new evaluation set of
76 pieces written by five professional composers, including for the first time newly composed
modern atonal works and rare coverage of instruments such as bassoon and viola. These instruments sparsely in existing datasets like MusicNet and URMP. The audio was rendered directly
from the MIDI scores using FluidSynth with the FluidR3 GM soundfont. Each composition
followed a set of predefined rules announced to all participants:
(1) Tempi restricted to 60--90 BPM;
(2) Meters limited to 3/4, 4/4, or 6/8;
(3) Maximum rhythmic subdivision of sixteenth notes;
(4) Richer notational elements such as swing, double-dotted notes, grace notes, and trills were excluded to reduce ambiguity in alignment;
(5) Pitch range limited to C2--C7 (five octaves);
(6) Dynamic markings restricted between pianissimo and fortissimo;
(7) Eight instruments were allowed (Piano, Violin, Viola, Cello, Flute, Bassoon, Trombone, Oboe), with at most three instruments per piece;
(8) At most one string instrument could appear in a piece.
These constraints were chosen to balance \emph{musical realism} (e.g., polyphony, dynamics, instrument variety) with \emph{evaluation clarity}, avoiding edge cases where subjective interpretation or notational ambiguity might dominate scoring. The constraints were also applied to simplify the task for the competitors, and to ensure the model's scoring remained sufficiently high.

\section{Competition Results}

A total of 21 teams registered for the 2025 Automatic Music Transcription Challenge, of which 14 teams submitted at least one solution. Eight teams submitted valid solutions whose MIDI outputs could be successfully graded without errors.
Table~\ref{tab:results} reports the results across all evaluated models, including open-source baseline (MT3). To protect privacy, teams can choose their own names. In the discussion, we focus on the top three systems, as lower-ranked submissions showed limited improvements over the baseline.

\begin{table}[h!]
    \centering
    \caption{Results from all evaluated systems. Multiple variants and public baselines are included; MT3 serves as the reference baseline. Runtime is measured by ms.
    }
    \label{tab:results}
    \begin{tabular}{@{}cp{1.5in}rrrrr@{}}
        \toprule
        Rank & Model Name                                                              & F1 Score        & Precision       & Recall          & Overlap         & Runtime       \\
        \midrule
        1    & MIROS                                                                   & \textbf{0.5998} & \textbf{0.6558} & 0.5724          & 0.7391          & 22.05         \\
        2    & YourMT3-YPTF-MoE-M                                                      & 0.5938          & 0.6010          & \textbf{0.5888} & 0.7305          & 12.60         \\
        3    & YourMT3-YPTF-S                                                          & 0.5581          & 0.5565          & 0.5615          & 0.7326          & 15.40         \\
        4    & YourMT3-P                                                               & 0.3947          & 0.3966          & 0.3985          & 0.7263          & 14.99         \\
        5    & MT3~\cite{gardner_mt3_2022} (baseline)                                  & 0.3932          & 0.3811          & 0.4115          & 0.7180          & 20.19         \\
        6    & YourMT3-YPTF-SP-V                                                       & 0.3305          & 0.3280          & 0.3358          & 0.7147          & 14.50         \\
        7    & press\_to\_win 1                                                        & 0.3199          & 0.3105          & 0.3346          & 0.7331          & 19.30         \\
        8    & press\_to\_win 2                                                        & 0.3190          & 0.3094          & 0.3331          & 0.7310          & 18.08         \\
        9    & YourMT3-YPTF-MoE-MP                                                     & 0.2173          & 0.2150          & 0.2206          & 0.6116          & 16.03         \\
        10   & press\_to\_win 3                                                        & 0.2168          & 0.2144          & 0.2203          & 0.6159          & 16.15         \\
        11   & Bytedance Piano~\cite{kong_high-resolution_2021}                        & 0.1721          & 0.2041          & 0.1689          & 0.5423          & 9.67          \\
        12   & press\_to\_win 4                                                        & 0.1470          & 0.1305          & 0.1799          & 0.6998          & 21.74         \\
        13   & ReconVAT~\cite{cheuk_reconvat_2021}                                     & 0.1415          & 0.1215          & 0.1803          & \textbf{0.7898} & 5.45          \\
        14   & Basic Pitch~\cite{2022_BittnerBRME_LightweightNoteTranscription_ICASSP} & 0.0634          & 0.0550          & 0.0782          & 0.5977          & \textbf{3.91} \\
        \bottomrule
    \end{tabular}
\end{table}

{\bf Music Information Retrieval Osnabrück (Winning System).}
Music Information Retrieval Osnabrück (MIROS) extends the YourMT3+ encoder–decoder framework for automatic music
transcription by adopting \emph{MusicFM} as the encoder backbone and pairing it with modernized decoders. MusicFM is a conformer-based, self-supervised
foundation model for music pretrained via BEST-RQ masked token modeling
\citep{won2024foundation}. An advantage of self-supervised foundation models is that they can exploit abundant unlabeled audio, compared with scarce labeled AMT data.
To integrate MusicFM into the YourMT3+ multi-decoder formulation,
MIROS introduced a recurrent adapter that conditions the temporally
downsampled encoder outputs on learned instrument group embeddings.
Each instrument group is then decoded in parallel using T5-style decoders
with cross attention, updated with rotary position embeddings and
hardware-optimized attention (FlashAttention) \citep{dao2023flashattention}.
Unlike prior YourMT3+ state-of-the-art systems, MIROS did not employ
cross-stem augmentation, though they did train on all available
MIDI datasets \citep{changyourmt3toolkittraining2022}.
This isolates the contribution of domain-pretrained audio representations within a comparable seq2seq framework.

Outside the competition, the MusicFM + multi-decoder system
($\approx$370M parameters) achieved a Slakh2100~\cite{manilow2019cutting} multi-instrument F-measure of 0.83,
with efficient long-context decoding (5-second windows, up to 1024 tokens per group)
and high training throughput ($\approx$2.4 iterations/s). Although it underperformed
YourMT3+ on Slakh2100, it attained slightly better accuracy on the competition data,
suggesting possible Slakh overfitting by YourMT3+. Domain-pretrained encoders like
MusicFM thus extends YourMT3+ to longer, richer contexts while maintaining
competitive accuracy, while optimized attention backends improve the practicality
of multi-instrument AMT at scale.

    {\bf General Trends Across Models.}
Several themes emerged from the technical directions of top models. Most submissions adopted a sequence-to-sequence (Seq2Seq) paradigm, extending the MT3~\cite{gardner_mt3_2022} architecture with enhancements such as self-supervised random projection quantizer~\cite{won2024foundation, chiu2022self}, Mixture-of-Experts~\cite{chang2024yourmt3+, jiang2024mixtral} routing, hierarchical time–frequency attention~\cite{lu2023multitrack}, cross-dataset augmentation~\cite{chang2024yourmt3+}, and auxiliary onset/offset losses. A second theme was efficiency: some teams targeted real-time transcription by pruning parameters, quantizing model weights, and lowering spectrogram resolution to reduce GPU memory and inference latency. Despite these advances, common failure cases were observed. Models often produced \emph{instrument leakage}, hallucinating nonexistent instruments, or struggled to disentangle salient melodies from dense polyphonic textures, especially when multiple instruments shared similar pitch ranges or timbres.


Since instrument leakage and polyphonic confusion emerged as common failure modes, we conducted a focused statistical analysis on transcription accuracy across exact instrument counts (1, 2, and 3 instruments). A two-way ANOVA on \texttt{f\_measure} with factors \emph{model} and \emph{instrument category} confirmed that the number of input instruments had a strong main effect ($F(2,219)=22.76$, $p<0.001$). Furthermore, the model main effect was also highly significant ($F(2,219)=14.49$, $p<0.001$), while the interaction remained non-significant ($F(4,219)=0.65$, $p=0.626$). This indicates that while the models differ significantly in their overall performance, they all suffer consistently as polyphony increases, regardless of architecture.

To expand this further, we conducted pairwise Welch's two-sample t-tests comparing excerpts with a single instrument against those with three instruments within each model. For the top system (MIROS), the degradation was large in magnitude ($t=3.11$, $p=0.0197$, Cohen's $d=1.40$). For YourMT3-YPTF-MoE-M, the difference was even more pronounced ($t=3.81$, $p=0.0103$, Cohen's $d=2.34$). Because multiple comparisons were performed across models, we also applied a Bonferroni correction. After correction, the difference for MIROS became borderline significant ($p\approx0.059$), while the YourMT3-YPTF-MoE-M result remained robustly significant ($p=0.031$). We report both raw and corrected $p$-values to distinguish statistical from practical significance. In both cases, the effect sizes exceeded $d=1.3$, which corresponds to very large practical differences as polyphony increases.

Table~\ref{tab:instrument_counts} provides descriptive statistics for the MT3 baseline and the top two systems, revealing a stark performance gap between single- and multi-instrument excerpts. While the models maintain high fidelity when transcribing solo passages, they struggle significantly to disentangle overlapping frequency content from multiple sources. For instance, when scaling from one to three instruments, the F-measure drops by over 0.28 points for MIROS and 0.36 points for YourMT3-YPTF-MoE-M. MIROS exhibits an especially sharp loss in precision, plummeting from 0.9067 on solo tracks to 0.4643 on three-instrument tracks. This consistent degradation highlights polyphonic transcription as the primary challenge for current approaches.

\begingroup
\setlength{\abovecaptionskip}{2pt}   
\setlength{\belowcaptionskip}{2pt}   
\renewcommand{\arraystretch}{0.92}   
\setlength{\tabcolsep}{4pt}          

\begin{table}[h!]
    \centering
    \caption{Performance comparison across exact instrument counts (1, 2, and 3 instruments) for the top two models alongside the MT3 baseline. Values are reported as mean $\pm$ standard deviation. Welch's t-test results are shown above in the text. The variable $n$ denotes the number of evaluated pieces belonging to each instrument count category.}
    \label{tab:instrument_counts}
    \begin{tabular}{@{}llccc@{}}
        \toprule
        Model              & Instrument Count & F-measure                    & Precision                    & Recall                       \\
        \midrule
        MT3 (Baseline)     & Count = 1 (n=6) & \textbf{0.5483 $\pm$ 0.3898} & \textbf{0.5548 $\pm$ 0.3974} & \textbf{0.5422 $\pm$ 0.3829} \\
        MT3 (Baseline)     & Count = 2 (n=24)& 0.3792 $\pm$ 0.1829          & 0.3661 $\pm$ 0.1808          & 0.3966 $\pm$ 0.1894          \\
        MT3 (Baseline)     & Count = 3 (n=46)& 0.2559 $\pm$ 0.1388          & 0.2475 $\pm$ 0.1409          & 0.2681 $\pm$ 0.1383          \\
        \midrule
        MIROS              & Count = 1 (n=6) & \textbf{0.7193 $\pm$ 0.2103} & \textbf{0.9067 $\pm$ 0.2286} & \textbf{0.6233 $\pm$ 0.2295} \\
        MIROS              & Count = 2 (n=24)& 0.5004 $\pm$ 0.2403          & 0.5460 $\pm$ 0.2296          & 0.4749 $\pm$ 0.2530          \\
        MIROS              & Count = 3 (n=46)& 0.4367 $\pm$ 0.2012          & 0.4643 $\pm$ 0.2044          & 0.4255 $\pm$ 0.2101          \\
        \midrule
        YourMT3-YPTF-MoE-M & Count = 1 (n=6) & \textbf{0.7594 $\pm$ 0.2304} & \textbf{0.7858 $\pm$ 0.2411} & \textbf{0.7367 $\pm$ 0.2254} \\
        YourMT3-YPTF-MoE-M & Count = 2 (n=24)& 0.4316 $\pm$ 0.2000          & 0.4436 $\pm$ 0.2043          & 0.4220 $\pm$ 0.1980          \\
        YourMT3-YPTF-MoE-M & Count = 3 (n=46)& 0.3918 $\pm$ 0.1471          & 0.3966 $\pm$ 0.1483          & 0.3882 $\pm$ 0.1465          \\
        \bottomrule
    \end{tabular}
\end{table}

\endgroup



Overall, nearly all top-performing models and the MT3 baseline relied on an almost identical set of ten datasets~\cite{gardner_mt3_2022, chang2024yourmt3+}. Notably, both YourMT3-P and MT3 applied cross-dataset augmentation on the same architecture and training corpus, yet achieved only marginal improvements. This suggests that performance is fundamentally constrained by data scarcity. In particular, currently available public datasets for AMT insufficiently cover less common instruments such as viola,  bassoon, and trombone, and are largely limited to a small number of ensemble pieces, thereby restricting generalization.

\section{Conclusion and Future Directions}
The 2025 Automatic Music Transcription Challenge advanced transcription of multi-instrument music by introducing a new test set and a cloud-based evaluation system. Several submissions surpassed the MT3 baseline, showing tangible gains but also exposing persistent weaknesses: dense polyphony, timbrally similar instruments, and limited data diversity. These limitations suggest that progress in AMT will depend as much on richer training resources as on architectural innovation.
Future iterations will expand to genres such as jazz and popular music and emphasize robust instrument detection. Additional improvements will target evaluation protocols that better capture pitch, timing, and timbre quality. By broadening coverage and refining benchmarks, future challenges aim to support the development of systems that generalize more reliably to diverse music.


\newpage
\bibliography{references, manual-reference}
\end{document}